\documentstyle[preprint,epsf,prb,aps]{revtex}

\newlength{\figwidth}
\begin{document}

\tighten
\draft

\title{A compact formula for the density of states in a quantum well}
\author{G. Iannaccone\cite{email} and B. Pellegrini}
\address{
Dipartimento di Ingegneria dell'Informazione:
Elettronica, Informatica e Telecomunicazioni \\
Universit\`a degli studi di Pisa, Via Diotisalvi 2,
I-56126 Pisa, Italy}
\date{To be published on Phys. Rev. B - cond-mat/9511012}
\maketitle

\begin{abstract}
In this paper we derive a formula for the density of states
in the presence of inelastic scattering
in the quantum well of a double barrier structure
as a function of a characteristic time
of the motion of electrons (namely, the round trip time in the well)
and of transmission probabilities for the whole structure and for each
barrier. In the model we use the scattering processes due to
phonons, impurities, and interface roughness, are taken into account
by a unique phenomenological parameter, the mean free path,
which plays the role of a relaxation length.
We also show that, for lower rates of incoherent processes,
the derived formula reduces to the one obtained by means of the
Breit-Wigner formalism.
\end{abstract}


\section{Introduction}

The density of states is one of the most important quantities for
the study of equilibrium and transport properties of quantum effect
devices. A recently derived formula \cite{ianna95} establishes
a simple general relation between the density of states in a
mesoscopic system and the dwell times for each incoming channel
connecting the system to the external world.

In this paper we apply that result to the quantum well of a double barrier
resonant tunneling diode. Moreover, in order to get closer to real systems
and experimental results, we consider the effects of inelastic processes
taking place in the well, which are not accounted for in the general relation
of Ref. \onlinecite{ianna95}.

We obtain a very compact formula connecting the density of states
in the well to a characteristic time of electron motion in double barrier
structures, i.e., the time an electron takes to complete a round trip of the
well, and to the transmission probabilities for the whole structure
and for each barrier.

In the literature the density of states in the well is usually obtained
using Breit-Wigner formulas,\cite{landau,weilvint87,buttiker88}
which lead to very simple and compact expressions. Dissipation can be easily
accounted for by introducing a partial resonance width for all the inelastic
processes.\cite{buttiker88,stonelee85,jonsgrin87}
However, for Breit-Wigner formulas to hold true, it is necessary that all
the partial resonance widths be much smaller than the separation between the
resonant energy levels, and between each level and
the top and bottom of the potential well. This
condition establishes an upper limit on the rate of incoherent processes
for the applicability of the Breit-Wigner formulas.

On the contrary, our model is valid even when coherence is completely
destroyed. Scattering with phonons, impurities,
and interface roughness is accounted for by a unique phenomenological
parameter $l$, the mean free path, a concept which
is well established in solid state physics.\cite{ashcroft76}

We assume that an electron traversing an infinitesimal length $dx$
of the one-dimensional device structure, experiences a collision
with probability $dx/l$, and that electrons emerge from
collisions with an equilibrium distribution function in
a state with completely random phase. This means, for instance,
that the square amplitude of a plane wave function
of wave vector $k$ attenuates exponentially as it propagates along the
$x$-axis, with a characteristic length equal to $l$. It is a situation
very similar to an electromagnetic wave propagating in a dissipative
medium. The difference is that the number of electrons has to be
conserved, therefore electrons which seem to have disappeared
have actually made a
transition to a different state, with a phase completely
uncorrelated, so that there is no
quantum interference between these electrons
and those that have not undergone an incoherent process.
In this model, all collision processes are effective in
randomizing phase and energy, and we do not make any
difference between the effects of elastic scattering
(due to impurities and interface roughness) and
inelastic scattering (with phonons, for instance).
A more sophisticated model should take into account
these differences, and, as a minor improvement,
phase randomization and energy relaxation could be
split using a different characteristic
length for each process.

B\"uttiker \cite{buttiker88} proposed a model for the
inclusion of incoherent processes which is similar to the one we use.
There is an inelastic scatterer in the
well modeled by an extra branch leading away from the
conductor to an extra reservoir, which does not draw net
current, but permits phase randomizing events. Anyway, such
a model is valid for very small differences between electrode
chemical potentials and/or when energy relaxation is
not accounted for. \cite{notina}
In the model by Kn\"abchen, \cite{knabchen92}
the inelastic scattering probability $\epsilon$ for an
electron traversing the well introduced by B\"uttiker is simply substituted by
$\exp(-w/l)$, where $w$ is the well width and $l$ the mean free path.

In our model scattering is spread over the whole region,
and not concentrated in a single point,
and any potential profile can be considered.
No addictional condition is required to obtain formula (\ref{dosw})
for the density of states. The hypothesis of smooth potential
in the well is imposed in order to obtain the compact formula
(\ref{doswbella}). The importance of the energy relaxation
mechanism will be shown elsewhere.\cite{iannadopo}

As we shall show in section III, we use in our formula
a characteristic time of
electron motion in the well defined on the basis of Larmor
times for transmission and reflection.
\cite{baz_larmor67,sokobask87,buettiker83,leavaers89,haugstov89,leavens90,iannpell94,iannpell_larmor94}
The tunneling time problem is the subject of a
long-standing controversy: while the dwell time\cite{buettiker83}
is widely accepted in the scientific community, there is no
consensus on the actual time spent
in a region by transmitted and reflected
particles,\cite{haugstov89,landmart94,barkbrou94} due to both
the fact that there is no operator for time in quantum mechanics
(therefore we cannot perform a direct measurement of time) and
that electrons do not follow actual trajectories in the Copenhagen
interpretation.\cite{leavens93}
The Larmor times are obtained as the result of an
indirect measurement: a weak perturbation is applied to the region of
interest (i.e., a magnetic field, a real potential, or an imaginary
potential) and some variation in the properties of transmitted and
reflected particles is measured (spin precession, phase rotation, or
particle absorption, respectively).
\cite{sokobask87,buettiker83,leavaers89,haugstov89,leavens90,iannpell94,iannpell_larmor94,mugabrou92}
What is controversial about Larmor times is the interpretation of such results
of an indirect measurement as the ``actual'' times spent in the
considered region. However, this point is not relevant to the aim of the
present work, where we are just interested in deriving a relation
between the Larmor times and the density of states in a quantum well.

Our paper is organized as follows: in section II we calculate
the transmission and reflection probabilities by using the
transfer matrix technique; in section III we derive a formula for
the density of states in the case of completely coherent transport.
A completely analogous formula which takes into account
the effects of dissipation is obtained in section IV, and is
shown to reduce down to the Breit-Wigner formulas for lower
rates of incoherent processes in sec. V. A summary ends the paper.

\section{Transmission and reflection probabilities for double
barriers}

Let us refer to the case sketched in Fig. 1: the
one-dimensional potential energy profile $V(x)$ defines the
first barrier $(a,0)$, the well region $(0,w)$, and the
second barrier $(w,b)$. Let us also introduce the wave vector $k(x)
\equiv [ 2 m(E-V(x))]^{1/2}/\hbar$ for all $x$ where
$V(x)<E$, where $m$ is the electron effective mass in the
material of the well, and $\hbar$ is the reduced Planck's
constant.

We can calculate the total transmission and reflection
coefficients by using the transfer matrix technique.
\cite{erdohern83,riccazbe84,ferry90,liusoll94}
In the assumption of coherent transmission through each single barrier,
the transfer matrix $M_1$ for the first barrier satisfies all the
properties listed in Ref. \onlinecite{erdohern83} and has
the form
\begin{equation}
M_1 = \left[ \begin{array}{cc}
1/t_1^{(l)} & r_1^{(l)*}/t_1^{(l)*} \\
r_1^{(l)}/t_1^{(l)} &1/t_1^{(l)*} \\
\end{array} \right]
, \label{tm1}
\end{equation}
where $t_1^{(l)}$ and $r_1^{(l)}$ are the
transmission and reflection coefficients, respectively, for
a plane wave coming from the left electrode, with wave vector
$k_{1l} \equiv k(a)$.  The corresponding coefficients $t_1^{(r)}$
and $r_1^{(r)}$ for a particle coming from the right with
wave vector $k_{1r} \equiv k(0)$ are $t_1^{(r)} = t_1^{(l)}
k_{1r}/k_{1l}$ and $r_1^{(r)} = - r_1^{(l)*} t_1^{(l)} /
t_1^{(l)*}$.
Moreover, if $T_1 = |t_1^{(l)}|^2
k_{1r}/k_{1l} = |t_1^{(r)}|^2 k_{1l}/k_{1r}$ is the
transmission probability, and $R_1 = |r_1^{(l)}|^2 =
|r_1^{(r)}|^2 $ is the reflection probability, we have $R_1
+T_1 = 1$, i.e., the continuity equation for the probability
density current holds true.
The same considerations apply to the second barrier and its
transfer matrix $M_2$, provided that we define
$k_{2l} \equiv k(w)$, $k_{2r} \equiv k(b)$ and change all
the subscripts $1$ into $2$.

In the well region, dissipative processes are accounted for
by means of the mean free path $l$; that is, the intensity of a
plane wave of wave vector $k'$
has a decay length equal to $l$.  As a consequence, the
probability density current for a given wave function
is not conserved. The effect of $l$ is taken in to account by
using a complex wave vector $k'_i = k' + i/2l$: a plane wave
of wave vector $k'_i$ along the $x$-axis has the form
$\exp(ik'_ix) = \exp(ik'x) \exp(-x/2l)$, therefore its square modulus
decays as $\exp(-x/l)$.

The multistep potential approximation\cite{andoitoh87}
can be used to obtain the transfer matrix $M_w$ of the well, provided that
the complex wave vector $k_i(x) = k(x) + i/2l$ is used at any point $x$
in the well.
If we make the hypothesis that the
potential varies smoothly enough that a semiclassical
approximation is valid, we obtain \cite{riccazbe84,ferry90,liusoll94}
\begin{equation}
M_w \approx  \left[
\begin{array}{cc} \gamma^{-1} & 0 \\ 0 & \gamma \end{array}
\right]
,\label{tmwell}
\end{equation}
where
\begin{equation}
\gamma = \exp \left\{i\int_0^w k_i(x) dx \right\}
.\label{gamma}
\end{equation}

The transfer matrix $M_{db}$ for the whole double barrier
structure is given by\cite{erdohern83,riccazbe84,ferry90,liusoll94}
\begin{equation}
M_{db} = M_1 M_w M_2
	= \left[ \begin{array}{ll}
1/t_{db}^{(l)} &-r_{db}^{(r)}/t_{db}^{(l)} \\
r_{db}^{(l)}/t_{db}^{(l)} &
	[ t_{db}^{(l)} -  r_{db}^{(l)} r_{db}^{(r)}/t_{db}^{(l)} ]
\end{array} \right]
\label{tmdb}
,\end{equation}
where $t_{db}^{(l)}$ and $r_{db}^{(l)}$ are
the transmission and reflection coefficients, respectively, for an electron
coming from the left, and $t_{db}^{(r)}$ and $r_{db}^{(r)}$
are the corresponding coefficients for an electron coming
from the right electrode.
Straightforward calculation yields \cite{liusoll94}
\begin{equation}
t_{db}^{(l)} = \frac{t_1^{(l)} t_2^{(l)} \gamma}{1 - c}
\label{tl}
\end{equation}
and
\begin{equation}
r_{db}^{(l)}
= r_1^{(l)} \frac{1-c/R_1}{1-c}
\label{rl}
,\end{equation}
where
\begin{equation}
c = r_1^{(r)} r_2^{(l)} \gamma^2
.\label{c}
\end{equation}
The expressions
for $t_{db}^{(r)}$ and $r_{db}^{(r)}$ can be easily obtained
from (\ref{tl}) and (\ref{rl})
by substituting the subscripts $l$ with $r$, $1$ with $2$,
and viceversa.
An electron coming from the left has a
probability $T_{db}^{(l)} = |t_{db}^{(l)}|^2 k_{2r}/k_{1l}$
of being transmitted and a probability
$R_{db}^{(l)} = |r_{db}^{(l)}|^2$ of being reflected; there is
also a fraction
$S_{db}^{(l)} = 1 - (T_{db}^{(l)} + R_{db}^{(l)})$ of electrons
which have been absorbed (i.e., have undergone incoherent processes).
In a completely analogous way we can
define the corresponding probabilities
$T_{db}^{(r)}$, $R_{db}^{(r)}$, and $S_{db}^{(r)}$ for an
electron coming from the right.
We also obtain that
$T_{db}^{(l)} = T_{db}^{(r)}$---hence we will often write
simply $T_{db}$---, while,
in general, $R_{db}^{(l)} \neq R_{db}^{(r)}$.

\section{Density of states in the case of coherent transport}

In this section we will address the case of no incoherent process
in the well, i.e., the mean free path $l \rightarrow \infty$. In
this case we have $R_{db}^{(l)} =  R_{db}^{(r)} = R_{db} = 1 - T_{db}$
and the
results of Ref. \onlinecite{ianna95} can be straightforwardly applied.

We showed\cite{ianna95} that the density of states in a given system is equal
to
the sum of dwell times corresponding to
each incoming channel divided by Planck's constant.
In our case the region $\Omega$ of interest
is $(a,b)$, and there are two incoming channels, the left and
the right ones, so that the density of states $\rho_\Omega(E)$
in $\Omega$, including both spin contributions, can be written as
\begin{equation}
\rho_\Omega(E) = \frac{1}{\pi\hbar}
	[ \tau_D^{(l)}(E) + \tau_D^{(r)}(E) ]
,\label{dos1}
\end{equation}
where $\tau_D^{(l)}$ and $\tau_D^{(r)}$ are the dwell times for an
electron of energy $E$ coming from the left and the right electrode,
respectively.

In order to obtain $\tau_D^{(l)}$ to substitute in
(\ref{dos1}), we can use the additivity of transmission and
reflection times $\tau_T^{(l)}$ and $\tau_R^{(l)}$ obtained
by using the Larmor clock and other well known approaches.
\cite{baz_larmor67,sokobask87,buettiker83,leavaers89,haugstov89,leavens90,iannpell94,iannpell_larmor94}
If we consider an electron coming from the left electrode, apply
a uniform perturbative potential $\lambda$ on the double barrier
($a < x < b$), and re-calculate the total transmission and
reflection coefficients as a function of $\lambda$, we can write
\begin{equation}
\tau_T^{(l)} \equiv  \mbox{Re} \left.
			\left\{ i \hbar
			\frac{\partial}{\partial \lambda} \ln t_{db}^{(l)}
			\right\}
			\right|_{\lambda = 0}
\label{tautl}
,\end{equation}
\begin{equation}
\tau_R^{(l)} \equiv  \mbox{Re} \left. \left\{ i \hbar
                        \frac{\partial}{\partial \lambda} \ln r_{db}^{(l)}
                        \right\}
			\right|_{\lambda =0}
\label{taurl}
,\end{equation}
and, finally, obtain
\cite{leavaers89,haugstov89,leavens90,iannpell94,iannpell_larmor94}
$\tau_D^{(l)} = \tau_T^{(l)} T_{db} + \tau_R^{(l)} R_{db} $.
In the Introduction we mentioned the controversy on the tunneling time
problem, and we are aware of the fact that there is no wide consensus
in the scientific community on the ``actual'' significance of
$\tau_T^{(l)}$ and $\tau_R^{(l)}$.
Anyway, re-assuring the reader that we do not want to forget about the long
debate in this field, for convenience reasons we will
refer to (\ref{tautl}) and (\ref{taurl}) as transmission and reflection
times.

Substitution of (\ref{tl}) and (\ref{rl}) in (\ref{tautl}-\ref{taurl}),
after straightforward but cumbersome calculations, yields
\begin{eqnarray}
\tau_D^{(l)} & = &
\tau_{T1}^{(l)} T_{db} +
\tau_{R1}^{(l)} R_{db} +
\tau_w T_{db} +
\tau_{T2}^{(l)} T_{db} \nonumber \\
& & +
\tau_{rt} T_1 \frac{|c|^2}{|1-c|^2}
,\label{taudl2}
\end{eqnarray}
where $\tau_{T1}^{(l)}$, $\tau_{T2}^{(l)}$, and $\tau_w$
are the transmission times for barriers 1 and 2, and the
well, respectively, defined as $\tau_T^{(l)}$ in (\ref{tautl})
replacing $t_{db}^{(l)}$
with $t_1^{(l)}$, $t_2^{(l)}$, $\gamma$, respectively;
$\tau_{R1}^{(l)}$ is the reflection time for barrier 1, defined
as $\tau_R^{(l)}$ in (\ref{taurl})
with $r_1^{(l)}$ in the place of $r_{db}^{(l)}$.
We call $\tau_{rt}$ {\em round trip time}; it is defined as
\begin{equation}
\tau_{rt} \equiv \mbox{Re} \left\{ \frac{i \hbar}{c}
\frac{\partial c}{\partial \lambda} \right\}
=
\tau_{R1}^{(r)} + 2 \tau_{w} + \tau_{R2}^{(l)}
\label{taurt}
;\end{equation}
the last equality derives from (\ref{c}) and
explains the name given to $\tau_{rt}$: it is actually the
sum of the times corresponding to the steps needed for a
round trip of the well: reflection from barrier 1,
traversal of the well, reflection from barrier 2, and again
traversal of the well.
We can easily obtain $\tau_D^{(r)}$ by repeating all the passages
from (\ref{tmdb}) to (\ref{taudl2}) commuting the subscripts
$l$ with $r$, and $1$ with $2$.
If we substitute (\ref{taudl2}) and the corresponding
result for $\tau_D^{(r)}$ into (\ref{dos1}) we obtain
\begin{eqnarray}
\rho_\Omega(E) & = & \frac{1}{\pi\hbar}
	[ (\tau_{T1}^{(l)} T_{db} + \tau_{R1}^{(l)} R_{db} )\nonumber \\
& + &	(\tau_{T2}^{(r)} T_{db} + \tau_{R2}^{(r)} R_{db}) \nonumber \\
&+& ( \tau_{T2}^{(l)} + 2 \tau_w + \tau_{T1}^{(r)} ) T_{db} \nonumber\\
&+& \tau_{rt} \frac{|c|^2}{|1-c|^2} ( T_1 + T_2 ) ]
\label{dos2}
;\end{eqnarray}
$\rho_\Omega(E)$ includes all the states in the region ($a<x<b$).
We are actually interested in the
states in the well region and in the tail states
penetrating both barriers on the well side, i.e., the states in the
``effective'' well region, therefore we drop from $\rho_\Omega$
the terms which take into account the states on the
left side of barrier 1 and on the right side of barrier 2 (i.e., the first
and the second term of (\ref{dos2}), respectively). Moreover,
the third term is easily shown to be much smaller (under the condition
$T_1,T_2 \ll 1$) than the fourth one, therefore the
density of states $\rho_w$ in the effective well region can be
written as
\begin{equation}
\rho_w(E) = \frac{1}{\pi \hbar}
\tau_{rt} \frac{|c|^2}{|1-c|^2} ( T_1 + T_2 )
.\label{doswell}
\end{equation}
{}From (\ref{c}) and from the fact that  $T_1,T_2 \ll 1$
we have $|c|^2(T_1 -T_2) \approx 1 -|c|^2$, so we get
\begin{equation}
\rho_w = \frac{1}{\pi\hbar} \tau_{rt} F(c)
,\label{doswell2}
\end{equation}
where we have defined
\begin{equation}
F(c) \equiv \frac{1 - |c|^2}{|1-c|^2}
.\label{fdic}
\end{equation}
The density of states in the effective well region is therefore
shown to be proportional to the round trip time times a factor
F(c), which will be shown in the next section
to depend only upon transmission and reflection
probabilities for the whole structure and for each barrier.

\section{Density of states in the presence of incoherent processes}

\subsection{Local density of states in the well}

In this section incoherent processes are taken into account, therefore
the formula (\ref{dos1}) for the density of states is no longer applicable.
The total density of states $\rho_w$ in the effective well region is obtained
as the sum of the density of tail states $\rho_1^{(r)}(E)$ and
$\rho_2^{(l)}(E)$
penetrating both barriers on the well sides
(Sec. IVB) and the integral of the local
density of states $\rho(x,E)$ in the well.

In this section,
in particular, we obtain a formula for $\rho(x,E)$
which does not require the hypothesis of smooth potential in the well.
We consider a  point $x'$ inside
the well ($x' \in (0,w)$). Let us split the $x$-axis
into two regions, and let us consider the potentials
$V_l(x) \equiv V(x)$ for $x<x'$ and $V(x')$ otherwise, and
$V_r(x) \equiv V(x)$ for $x>x'$ and $V(x')$ otherwise, as sketched
in Fig. 2. Let us call $r_l(x')$ the reflection coefficient for a plane wave
of energy $E$ incident on $V_l(x)$ from the right, and
$r_r(x')$ the reflection coefficient for
a plane wave of energy $E$ impinging on $V_r(x)$ from the left.
The local density of states at a point $x'$
can easily be written as \cite{ianna95}
\begin{equation}
\rho(x',E) = \frac{1}{\pi \hbar} \sum_{n=1}^2 \frac{|\psi_n(x',E)|^2}{J_n(E)}
\label{ldos}
,\end{equation}
where both spin contributions have been considered,
the wave functions are not normalized and $J_n(E)$ is the total
current associated to state $\psi_n$ entering the whole system.
The sum is over all degenerate states corresponding to the same
energy $E$, i.e., in our case, the ones associated to a particle
coming from the left electrode ($n=1$) and a particle coming from
the right electrode ($n=2$). The quantities
to be put in (\ref{ldos}) are derived in the Appendix. Substitution of
(\ref{A1}-\ref{A2}) and the corresponding quantities for $\psi_2$ in
(\ref{ldos}) yields:
\begin{equation}
\rho(x',E)=\frac{1}{\pi \hbar} \mbox{Re} \left \{
	\frac{ (1 + r_r(x'))(1 + r_l(x')) }
	     { 1 - r_r(x') r_l(x') }
	\right \}
.\label{ldos2}
\end{equation}
An identical result has been obtained through different procedures and
in simplified conditions by other authors. \cite{knabchen92,modinos}

\subsection{Density of tail states penetrating both barriers on the well sides}

For obtaining the density of tail states penetrating both
barriers on the well sides
we just have to use (\ref{ldos}), provided
that the sum is only over the states incident on the well side of the
barriers.
Let us consider the first barrier: the density of tail states
$\rho_1^{(r)}$ is
\begin{equation}
\rho_1^{(r)}(E) = \frac{1}{ \pi \hbar}
		\sum_{n=1}^2
		\frac{\int_a^0 |\psi_n(x,E)|^2 dx}{J_n(E)}
.\label{dos1r}
\end{equation}
We can find for $\rho_1^{(r)}$ a more compact expression:
in the Appendix we derived the currents $J_{1inc}^{(r)}(x,E)$ and
$J_{2inc}^r(x,E)$ incident on $V_l(x)$, associated to the states $\psi_1$ and
$\psi_2$, respectively. Therefore, if we remember
that the dwell time is defined as the ratio of the
integral of the probability density over the considered region to the
incident current, from (\ref{A3}-\ref{A4}) we can write
\begin{eqnarray}
\rho_1^{(r)}(E) & = &
	\frac{1}{\pi \hbar}
	\left[
	\frac{J_{1inc}^{(r)}(0,E)}{J_1(E)}
	+
	\frac{J_{2inc}^{(r)}(0,E)}{J_2(E)}
	\right]
	\tau_{D1}^{(r)}
	\nonumber \\
& = & \frac{1}{\pi \hbar} \tau_{D1}^{(r)} F[r_r(0) r_l(0)]
,\label{dos1rbis}
\end{eqnarray}
where the function $F$ has been already defined in (\ref{fdic}).

For the second barrier, following the same procedure, we have
\begin{equation}
\rho_2^{(l)}(E) = \frac{1}{\pi \hbar} \tau_{D2}^{(l)} F[r_r(w) r_l(w)]
.\label{dos2l}
\end{equation}

\subsection{Density of states in the effective well region}

The density of states in the effective well region is therefore, from
(\ref{ldos2}),
(\ref{dos1rbis}), and (\ref{dos2l}),
\begin{equation}
\rho_w(E) = \rho_1^{(r)}(E) + \rho_2^{(l)}(E) + \int_0^w \rho(x,E) dx
\label{dosw}
.\end{equation}

We can write $\rho(x,E)$ in a different way, in order to derive a
more compact formula for $\rho_w(E)$. Let the density
matrix $\hat{g}(E)$ in the well be the incoherent superposition
of states $\psi_1$ and $\psi_2$ with probabilities $p_1$
and $p_2$, i.e.,
\begin{equation}
\langle x | \hat{g}(E) | x \rangle = p_1 | \psi_1(x)|^2 +
				p_2 | \psi_2(x)|^2
\label{gdie}
.\end{equation}
Associated to $\hat{g}(E)$ there is the probability density current $J(E)$,
whose expression is given by (\ref{A1}) and the corresponding
quantity for $\psi_2$,
that can be split into a left going component $J_l(x,E) = p_1
J_{1inc}^{(l)}(x,E)
+ p_2 J_{2inc}^{(l)}(x,E)$ and a right going component
$J_r(x,E) = p_1 J_{1inc}^{(r)}(x,E) + p_2 J_{2inc}^{(r)}(x,E)$.
Now, we can make the hypothesis that both $J_l$ and $J_r$ are much
greater than the net current $J = J_l -J_r$. By imposing $J_r \approx
J_l$ we can obtain $J_r$, $J_l$, and $\langle x | \hat{g}(E) | x \rangle$
as a function of $p_1/p_2$.
This result, substituted in (\ref{ldos2}), yields
\begin{equation}
\rho(x',E) = \frac{1}{\pi \hbar}
		\frac{\langle x | \hat{g}(E) | x \rangle}{J_l(x',E) + J_r(x',E)}
		F[r_l(x') r_r(x')]
.\end{equation}

A great simplification of (\ref{dosw}) can be obtained if we make
again the semiclassical approximation in the well. In fact, in this
case we have
just to notice that
$r_r(x') r_l(x') = c$ for all $x \in (0,w)$. Therefore we can write
\begin{equation}
\rho_w(E) = \frac{1}{\pi \hbar}
		\tau_{rt}^i
		F(c)
\label{doswfinale}
\end{equation}
where we have defined
\begin{equation}
\tau_{rt}^i \equiv \tau_{D1}^{(r)} + \tau_{D2}^{(l)} + 2 \int_0^w
		\frac{ \langle x' | \hat{g}(E) | x' \rangle }
			{ J_l(x') + J_r(x') } dx'
;\label{taurti}
\end{equation}
$\tau_{rt}^i$ can be interpreted as the round trip time of the
well in the presence of
inelastic processes. We wish to point out that (\ref{doswfinale})
is formally analogous to (\ref{doswell2}) found in the case
of coherent transport. It is also easy to verify that
when $l \rightarrow \infty$, i.e., when the
limit of coherent transport is approached, $\tau_{rt}^i$ tends to
the value of $\tau_{rt}$ defined in (\ref{taurt}).

The condition $T_1,T_2 \ll 1$ allows us to write, with
very good approximation, the following expression for $F(c)$,
where $T_{db}$, $R_{db}^{(r)}$, and $R_{db}^{(l)}$ are
obtained from (\ref{tl}) and (\ref{rl}):
\begin{equation}
F(c) \approx \frac{1-R_{db}^{(l)}}{T_1} + \frac{T_{db}}{T_2}
\approx \frac{1-R_{db}^{(r)}}{T_2} + \frac{T_{db}}{T_1}
;\label{fdicfinale}
\end{equation}
therefore (\ref{doswfinale}) can be written as
\begin{equation}
\rho_w(E) \approx \frac{1}{\pi \hbar}
                \tau_{rt}^i
\left[\frac{1-R_{db}^{(l)}}{T_1} + \frac{T_{db}}{T_2}  \right]
,\label{doswbella}
\end{equation}
i.e., as a function of the round trip time and transmission and
reflection probabilities for the total structure and for each
barrier.

\section{Comparison with Breit-Wigner formulas}

In this section we want to show that for a rate
of incoherent processes low enough (i.e., a long
enough mean free path), the formula for the density of states
derived above reduces to the one obtained by the means
of Breit-Wigner formulas.

Let us expand $c$ given by (\ref{c}) to first order around
the resonant energy $E_R$
(which is the energy at which $c$ is real and positive):
\begin{equation}
c(E) \approx c(E_R) \left[ 1 - i \frac{\tau_{rt}^i(E_R)}{\hbar}(E-E_R) \right]
\label{cetaylor}
\end{equation}
where we have used the definition (\ref{taurt}) of $\tau_{rt}$ and the
fact that $\arg \{ i \hbar \partial \log c /\partial E \} \approx 0$.
If $w/l$ is small enough we can write
\begin{eqnarray}
c(E_R) & = &[R_1(E_R) R_2(E_R)]^{1/2} e^{-w/l} \nonumber \\
& \approx & 1 - \frac{1}{2}T_1(E_R) - \frac{1}{2}T_2(E_R) - \frac{w}{l}
\label{cer}
.\end{eqnarray}

Substitution of (\ref{cetaylor}) and (\ref{cer}) into (\ref{tl})
and (\ref{rl}) yields
\begin{equation}
T_{db} \approx \frac{\Gamma_1 \Gamma_2}
        {(E-E_R)^2 + (\Gamma/2)^2}
\label{tdbbreit}
\end{equation}
and
\begin{equation}
1 - R_{db}^{(l)} \approx \frac{ \Gamma_1 (\Gamma_2 + \Gamma_i)}
                        {(E-E_R)^2 + (\Gamma/2)^2}
,\label{rdblbreit}
\end{equation}
where
$\Gamma_1 = \hbar T_1(E_R) /\tau_{rt}^i(E_R)$,
$\Gamma_2 = \hbar T_2(E_R) /\tau_{rt}^i(E_R)$,
$\Gamma_i = \hbar 2 w /[ \tau_{rt}^i(E_R) l]$, and
$\Gamma = \Gamma_1 + \Gamma_2 + \Gamma_i$.

Equations (\ref{tdbbreit}) and (\ref{rdblbreit}) are
the Breit-Wigner formulas,\cite{landau,buttiker88,garcrubi89} and
$\Gamma_1$, $\Gamma_2$, and $\Gamma_i$ are the partial resonance widths for
each process allowing escape from the resonant state, in
particular tunneling through barrier 1 and 2, and incoherent processes,
respectively. Partial resonant widths are characteristic quantities
of the Breit-Wigner formalism, and are given by the ratio between $\hbar$ and
the characteristic time of the process we are considering.
In the case of escape through one of the barriers the
characteristic time is intuitively given by the ratio of the round trip
time and the tunneling probability of the barrier.
In the case of inelastic scattering the time is $\tau_{rt}$ times
the ratio between the mean free path $l$ and the length corresponding
to a round trip of the well ($2w$).

{}From (\ref{doswbella}) and (\ref{tdbbreit}-\ref{rdblbreit}),
we straightforwardly have
\begin{equation}
\rho_w(E) \approx \frac{1}{\pi} \frac{\Gamma}{(E-E_R)^2 +
(\Gamma/2)^2}
,\end{equation}
i.e., the result usually obtained from Breit-Wigner formulas.
\cite{weilvint87,buttiker88}
We wish to point out that this formula holds true if
the development of $c$ to first order of $E-E_R$
and to first order in $w/l$ is a good approximation. In other words,
Breit-Wigner formulas can be used if each partial
width is much smaller than both the resonant energy $E_R$
and the difference between the height of the barriers and $E_R$.
In our case these conditions are true for $\Gamma_1$ and $\Gamma_2$
and holds true for $\Gamma_2$ if $l/w$ is high enough.
Otherwise, the expression given by (\ref{doswfinale}), which has a wider
range of applicability, has to be used.

\section{Summary}

In this paper we have studied the density of states
in a double barrier structure. We have proposed a simple
model which is able to account for inelastic processes
occuring in the quantum well by means of a single
phenomenological parameter $l$, the mean free path.

We have obtained a very compact formula which relates
the density of states in the effective well region
to the round trip time of the
quantum well and to the tunneling probabilities for
the single barriers and for the whole structure.
The formula is valid both in the case of completely
coherent transport and in the case when dissipative
processes in the well are predominant.

This formula will be shown to be fundamental in
unifying two widely known description of transport
in double barrier structures: that of resonant tunneling
and the one of sequential tunneling. \cite{iannadopo}

We believe that the role of the density of states---a
characteristic quantity of a system in equilibrium---in
the steady state transport and in the characteristic
times of the motion of electrons in a mesoscopic system
deserves a deeper investigation.

\section{Acknowledgments}

The present work has been supported by the Ministry for the University
and Scientific and Technological Research of Italy, by the Italian
National Research
Council (CNR).

\appendix

\section*{}

Let us consider a point $x=x'$ in the well and the state $\psi_1$
of energy $E$ corresponding to a particle coming from the left of $x'$. We
can describe $\psi_1$ as a plane wave of amplitude 1 undergoing
multiple reflections on $V_r(x)$ for $x>x'$ and $V_l(x)$ for
$x<x'$ (see Figs. 1 and 2), so that we have
\begin{equation}
\psi_1(x') = 1 + r_r(x') + r_r(x') r_l(x') + \cdots
	= \frac{1 + r_r(x')}{1 - r_r(x') r_l(x')}
{}.
\label{A1}
\end{equation}
Now, $[1 - |r_l(x')|^2]$ is the probability that a particle impinging
on $V_l(x)$ is not reflected back (i.e., is either transmitted or
``absorbed'' on the left of $x'$). For time reversal symmetry, it is
also the probability that an electron coming from the left of $x'$
appears at $x=x'$: if 1 is the amplitude of $\psi_1$ before taking into
account multiple reflections, the total current entering the system
has to be
\begin{equation}
J_1 = \frac{v(x')}{1 - |r_l(x')|^2}
,\label{A2}
\end{equation}
where $v(x) = \hbar k(x)/m$.
It is worthy noticing that the dependence of $J_1$ on $x'$ is due only to
the fact that $\psi_1$ is not normalized.
We can also  associate to $\psi_1$ and $x'$ a probability current density
which can be split into a left going component $J_{1inc}^{(l)}$ (incident on
$V_l(x)$), and a right going component $J_{1inc}^{(r)}$ (incident on
$V_r(x)$):
\begin{equation}
J_{1inc}^{(l)}(x') = \frac{|r_r(x')|^2 v(x')}{|1 - r_l(x')r_r(x')|^2}
\label{A3}
\end{equation}
and
\begin{equation}
J_{1inc}^{(r)}(x') = \frac{v(x')}{|1 - r_l(x')r_r(x')|^2}.
\label{A4}
\end{equation}
The corresponding quantities for the state $\psi_2$, associated to a particle
coming from the right of $x=x'$, can be obtained by substitution of 1 with 2,
$r$ with $l$, and viceversa.

\begin{figure}

\vspace{2cm}

\epsfxsize=\figwidth
\epsffile{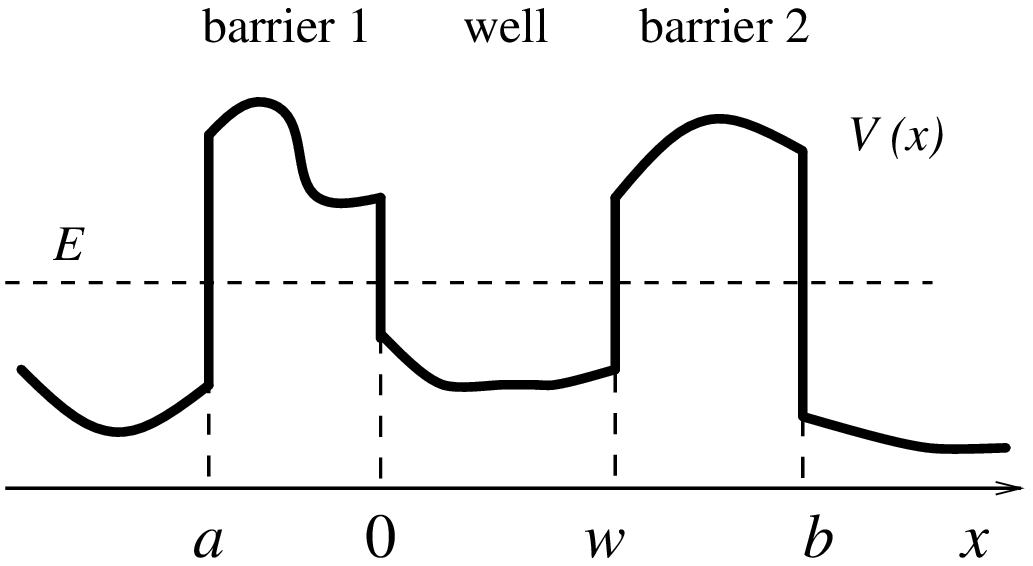}

\vspace{1cm}

\caption{The one-dimensional potential $V(x)$ defines the two barriers [$(a,0)$
and
$(w,b)$] and the well $(0,w)$.}
\end{figure}

\begin{figure}

\epsfxsize=\figwidth
\epsffile{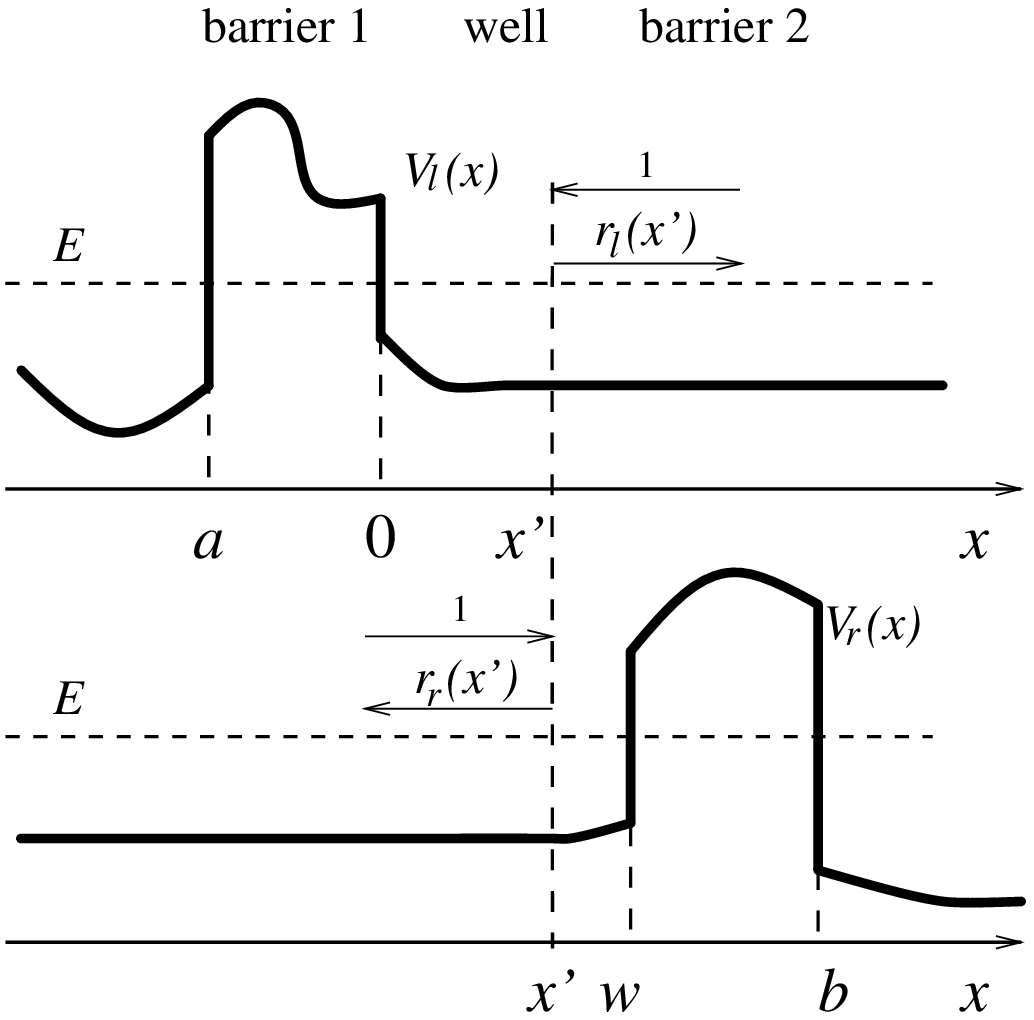}

\vspace{1cm}

\caption{The double barrier structure is split into two regions at $x=x'$:
$r_r(x')$ is the reflection coefficient for a plane wave of energy
$E$ incident on $V_r(x)$ from the left and $r_l(x')$ is the reflection
coeffincient
for a plane wave of energy $E$ incident on $V_l(x)$ from the right.}
\end{figure}

\begin{references}
\bibitem{notina}
Eq. A4-A12 of Ref. 4 imply that transmission and reflection
probabilities do not depend upon energy, i.e., that electrode
chemical potentials are close enough that electrons contributing
to the net current have energies in a very narrow energy range.
Otherwise, the incoherent transmission probability in A12 of Ref. 4
is still valid, provided that we do not include energy relaxation
in the extra reservoir and impose on the extra branch that all
current contributions at any energy be zero, and not only the net
current flow.
\end{references}
\end{document}